\documentclass[10pt]{article}

\usepackage{amstex}

\def\u{\underline}
\def\o{\omega}
\def\d{\displaystyle}
\def\phi{\varphi}
\def\I{\mathbb{I}^-}
\def\Pa{\u{\partial}}

\begin{document}

\noindent{\Large \bf
KAM--renormalization-group for Hamiltonian systems with two degrees of freedom
\normalsize \rm}

\vspace{24pt}

\noindent{\bf  C.\ Chandre and H.R.\ Jauslin}

\vspace{12pt}
\it
\noindent
Laboratoire de Physique-CNRS, Universit\'e de Bourgogne\\
B.P.\ 400, F-21011 Dijon, France
\rm

\vspace{24pt}
\footnotesize
\begin{quote}
{\bf Abstract:} We review a formulation
of a renormalization-group scheme for Hamiltonian
systems with two degrees of freedom. 
We discuss the renormalization flow on the basis of
the continued fraction expansion of the frequency.
The goal of this approach is to understand universal scaling behavior
of critical invariant tori.

\vspace{3pt}


\end{quote}
\normalsize

\vspace{12pt}

\section{Introduction}
\setcounter{equation}{0}

Recently, renormalization-group ideas have been proposed to describe the 
breakup of invariant tori for Hamiltonian systems with two degrees of 
freedom~\cite{koch,govin,chandre,cgjk,abad}. Most of the numerical work
has been done for the golden mean torus (mainly for practical convenience).
In this article, we extend the renormalization-group transformation to more
general frequencies.\\
We consider the following class of Hamiltonians with two degrees of
freedom, quadratic in the actions $\u{A}=(A_1,A_2)$:
$$
H(\u{A},\u{\phi})=\frac{1}{2}m(\u{\phi})(\u{\Omega}\cdot\u{A})^2
+[\u{\o}_0+g(\u{\phi})\u{\Omega}]\cdot \u{A} +f(\u{\phi}),
$$
where $m$, $g$, and $f$ are even scalar functions of the angles $\u{\phi}=
(\phi_1,\phi_2)\in \mathbb{T}^2=[0,2\pi)^2$; 
the frequency vector of the considered
torus $\u{\o}_0=(\o,-1)$ satisfies a Diophantine condition 
$|\u{\o}_0 \cdot \u{\nu}|\geq \sigma |\u{\nu}|^{-\tau}$ for all $\u{\nu}
\in \mathbb{Z}^2$, and for some $\sigma> 0$ and $\tau> 1$.
The rank of the matrix $\partial^2 H/ \partial \u{A}^2$ is equal to one
(this matrix is proportional to the projection operator on the
$\u{\Omega}$-direction).
Thus, $\mbox{det }[\partial^2 H/ \partial \u{A}^2]=0$, i.e.,
the Hamiltonians we consider do not satisfy the twist
condition. There is however a twist in a particular direction of the actions
characterized by the vector $\u{\Omega}=(1,\alpha)$, which is what is
needed to obtain KAM stability of Diophantine tori for small
perturbations~\cite{chandrejauslin}.\\
If one takes $g$ and $f$ equal to zero, the equations of motion are
$$
\frac{d\u{A}}{dt}=-\frac{1}{2}\frac{\partial m}{\partial \u{\phi}}(\u{\Omega}
\cdot \u{A})^2 \, , \qquad \frac{d\u{\phi}}{dt}=m(\u{\phi})(\u{\Omega}
\cdot \u{A})\u{\Omega} +\u{\o}_0.
$$
Then $\u{A}=\u{0}$ defines an invariant torus, and the motion on this torus
is quasiperiodic with frequency vector $\u{\o}_0$. For $g$ and $f$ sufficiently
small, a KAM theorem~\cite{chandrejauslin} proves the persistence
of this torus for arbitrary $m$, provided that its mean value is nonzero.
It is not necessary to have $m$ small in order to prove the 
existence of the torus with frequency vector $\u{\o}_0$. This remark allows 
us to perform canonical transformations that stay within
the class of Hamiltonians {\em quadratic} in the actions.\\
As one increases the perturbation (consisting of $g$ and $f$), the torus 
gets deformed. This deformation is related to families of
nearby periodic orbits that accumulate at the torus. This accumulation
motivates the setup of a renormalization transformation combining a 
{\em rescaling} of 
phase space and an {\em elimination} of the irrelevant part at each scale.\\
An attractive ({\em trivial}) fixed point of the renormalization
represents the phase where the torus exists.
An hyperbolic ({\em nontrivial}) fixed point 
(or more generally an hyperbolic fixed set) corresponds to a
transition where the torus breaks up.\\
For the golden mean torus with frequency vector $\u{\o}_0=(\gamma^{-1},-1)$
where $\gamma=(1+\sqrt{5})/2$, numerical studies suggest that the critical
surface is the stable manifold (of codimension 1) of a nontrivial fixed
point. Also for quadratic irrationals (those with periodic continued
fraction expansion), one expects a nontrivial fixed point for
a certain renormalization operator. For 
nonperiodic continued fraction expansion, hyperbolic fixed sets instead of
fixed points are 
conjectured~\cite{ostlund,lanford,lanford2,rand,umberger,satija}.

\section{Renormalization transformation}
\setcounter{equation}{1}

We describe the renormalization scheme for a torus with arbitrary
frequency vector $\u{\o}_0=(\o,-1)$ where $\o \in ]0,1[$.
This renormalization relies upon the continued fraction expansion of 
$\o$:
$$ \o=\frac{1}{a_0+\d\frac{1}{a_1+\cdots}} \equiv [a_0,a_1,\ldots].$$
The best rational approximants are given by the truncations of this 
expansion $p_k/q_k=[a_0,a_1,\ldots,a_k=\infty]$. The
corresponding periodic orbits with frequency vectors
$\u{\nu}_k=(q_k,p_k)$ (called ``resonances''
in what follows) accumulate at the invariant torus. This family of frequency
vectors satisfies
$|\u{\o}_0\cdot\u{\nu}_{k+1}| < |\u{\o}_0\cdot\u{\nu}_k|$ and
$
\lim_{k \infty} |\u{\o}_0\cdot\u{\nu}_k|=0,
$
as it can be seen from the relation
$$
\u{\nu}_k=N_{a_0}\cdots N_{a_{k-1}}\u{\nu}_0,
$$
where $\u{\nu}_0=(1,0)$ and $N_{a_i}$ denotes the matrix
$$ N_{a_i}=\left( \begin{array}{cc} a_i & 1 \\ 1 & 0 \end{array} \right).$$
Moreover, $\u{\o}_0\cdot\u{\nu}_k$ and $\u{\o}_0\cdot\u{\nu}_{k+1}$ are of
opposite sign (as the stable eigenvalue of $N_{a_i}$ is negative); thus
the torus is approached from above and from below by the sequence of periodic
orbits with frequency vectors $\{ \u{\nu}_k \}$.\\
We notice that if the continued fraction expansion
is periodic with period $s$, then one can extract from the sequence
$\{\u{\nu}_k\}$, $s$ families of periodic orbits which accumulate
{\em geometrically}, with the same ratio, at the torus.\\
The word resonance refers to the fact that the small denominators
$\u{\o}_0\cdot\u{\nu}_k$ that appear in the perturbation expansion
are the smallest ones, i.e., $|\u{\o}_0\cdot\u{\nu}_k|<|\u{\o}_0\cdot\u{\nu}|$
for any $\u{\nu}=(q,p)$ different from zero and $\u{\nu}_k$, and such that
$|q|<q_{k+1}$.\\
The main scale of the perturbation
is defined by $\u{\nu}_0$ for the torus with
frequency $\o$. We will denote this scale by $[\u{\nu}_0,\o]$. The
next smaller scale is $[\u{\nu}_1,\o]$. The renormalization transformation 
changes the coordinates such that the next smaller scale becomes the main one,
i.e., the main scale is now $[\u{\nu}_0,\o']$ where $\o'=[a_1,a_2,\ldots]$.
As the frequency is changed (the continued fraction expansion is shifted 
to the left), the sequence of resonances is mapped into the
sequence
$$
\u{\nu}_k'=N_{a_1}\cdots N_{a_{k-1}}\u{\nu}_0.
$$
The renormalization transformation is a map $(m,g,f,\o,\alpha) \mapsto
(m',g',f',\o',\alpha')$, where $\alpha$ denotes the second component
of $\u{\Omega}=(1,\alpha)$.\\
The transformation $\mathcal{R}_{a_0}$ consists of four steps:\\
{\bf (1)} A shift of the resonances constructed by the
condition $\u{\nu}_1 \mapsto \u{\nu}_0$:
we require that $\cos[(a_0,1)\cdot \u{\phi}]=\cos [(1,0)\cdot\u{\phi}']$.
This change 
is done via a linear canonical transformation $(\u{A},\u{\phi})\mapsto 
(N_{a_0}^{-1}\u{A},N_{a_0}\u{\phi})$.
This step changes the frequency vector $\u{\o}_0$ into $\u{\o}_0'=(\o',-1)$,
since $N_{a_0}\u{\o}_0=-\o\u{\o}_0'$.\\
{\bf (2)} We rescale the energy (or time) by a factor $\o^{-1}$, and we change 
the sign of both phase space coordinates
$(\u{A},\u{\phi})\mapsto (-\u{A},-\u{\phi})$, in
order to have $\u{\o}_0'$ as the new frequency vector, i.e., the 
average of the term linear in the actions is of the form
$\u{\o}_0'\cdot\u{A}$. Furthermore,
$\u{\Omega}=(1,\alpha)$ is changed into $\u{\Omega}'=(1,(a_0+\alpha)^{-1})$.\\
{\bf (3)} Then we perform a rescaling of the actions: $H$ is changed into
$\hat{H}(\u{A},\u{\phi})=\lambda H(\u{A}/\lambda,\u{\phi})$ with 
$\lambda=\lambda(H)$ such that the mean value of $m$ is equal to 1, i.e.,
$\lambda=\o^{-1}(a_0+\alpha)^2\langle m \rangle$.\\
{\bf (4)} A canonical transformation that eliminates the nonresonant part of 
$g$ and $f$.\\
The choice of which part of the perturbation is resonant or not is somewhat 
arbitrary. Recall that what is relevant is the accumulation of resonances;
therefore a possible choice of irrelevant modes could include all modes 
except the resonances. However, it is desirable from a numerical point 
of view not to eliminate too many modes. A reasonable choice is the set
$\mathcal{C}$ of integer vectors $\u{\nu}$ such that $|\nu_2|> |\nu_1|$.
We notice that the relation defining $\u{\nu}_k=(q_k,p_k)$ shows that
$q_k\geq p_k$ for $k\geq 0$, i.e., the resonances are not elements of
$\mathcal{C}$.\\
From the form of the eigenvectors of $N_{a_0}$, one can see that every
$\u{\nu}\in \mathbb{Z}^2\setminus (0,0)$ goes into $\mathcal{C}$ after 
sufficiently
many iterations of matrices $N_{a_i}$ (as the unstable eigenvector
of $N_{a_0}^{-1}$, which is $\u{\o}_0$, points into $\mathcal{C}$).
In other terms, a resonant mode at some scale turns out to be a 
nonresonant one at a sufficiently smaller scale. We notice that $(0,0)$
is not an element of $\mathcal{C}$, i.e., it is resonant.\\
We eliminate completely all the nonresonant modes of $g$ and $f$ by a 
canonical transformation, connected to the identity, which is defined
by iterating KAM-type transformations.\\
We denote $\I$ the projection
operator on the nonresonant part:
$$\I f(\u{\phi})=\sum_{\u{\nu}\in \mathcal{C}}f_\nu e^{i\nu\cdot\phi},$$
and $\Pa$ the derivative with respect to the angles $\u{\phi}$:
$\Pa=\partial/\partial \u{\phi}$.\\
The KAM iterations we perform are generated by functions linear in the actions, 
and
it allows us, following Thirring~\cite{thirring}, to remain quadratic in the
actions at each step.\\
This can be seen by working with Lie transformations $\mathcal{U}_S:
(\u{A},\u{\phi})\mapsto (\u{A}',\u{\phi}')$ generated by 
$$
S(\u{A},\u{\phi})=Y(\u{\phi})\u{\Omega}\cdot\u{A}+Z(\u{\phi})
+a\u{\Omega}\cdot\u{\phi},
$$
characterized by two scalar functions $Y$ and $Z$, and a constant $a$.
The expression of the Hamiltonian in the new variables is obtained by 
\begin{eqnarray*}
H'&=&H\circ U_S= e^{\hat{S}}H \\
&=& H+\{ S,H\} +\{S,\{ S,H\}\}/2!+\cdots,
\end{eqnarray*}
where $\{\, , \, \}$ is the Poisson bracket of two functions of the 
action and angle variables
$$
\{S,H\}=\frac{\partial S}{\partial \u{\phi}}\cdot 
        \frac{\partial H}{\partial \u{A}}
       -\frac{\partial S}{\partial \u{A}}\cdot 
        \frac{\partial H}{\partial \u{\phi}}.
$$
From this equation, one can see that $H'$ is again quadratic in the actions:
$\partial S/\partial \u{\phi}$ and $\partial H/\partial \u{A}$ are linear 
in the actions, $\partial S/\partial \u{A}$ is action-independent and
$\partial H/\partial \u{\phi}$ is quadratic; therefore $\{S,H\}$ is quadratic
in the actions, and in fact, quadratic in the $\u{\Omega}\cdot\u{A}$
variable.\\
The generating function is determined such that it eliminates the nonresonant 
modes of $g$ and $f$. This cannot be defined directly, so one iterates an
infinite number of steps of such transformations such that one 
iteration eliminates
the order $\varepsilon$ while producing an order $\varepsilon^2$.
Following Ref.\ \cite{cgjk}, we have the equations
\begin{eqnarray*}
&& \u{\o}_0\cdot \Pa Z +\I f=const,\\
&& \u{\o}_0\cdot \Pa Y +\I g+\I (
m\u{\Omega}\cdot \Pa Z )+a\Omega^2 \I m=0, \\
&& \langle m \rangle a\Omega^2+\langle g\rangle+\langle m \u{\Omega}\cdot
\Pa Z \rangle=0.
\end{eqnarray*}
The constant $a$ corresponds to a translation in the action variables 
which has the purpose of eliminating the mean value of the linear term 
in the variable $\u{\Omega}\cdot \u{A}$.
We notice that $Y$ and $Z$ only contain nonresonant modes, e.g., $Y=\I Y$, and
that $Z$ (resp.\ $Y$) is chosen to reduce
$\I f$ (resp.\ $\I g$) from $\varepsilon$ to $\varepsilon^2$.
These equations are solved by Fourier series:
\begin{eqnarray*}
&&Z(\u{\phi})=\sum_{\u{\nu}\in\mathcal{C}}\frac{i}{\u{\o}_0\cdot\u{\nu}}
f_{\u{\nu}}e^{i\u{\nu}\cdot\u{\phi}}, \\
&&Y(\u{\phi})=\sum_{\u{\nu}\in\mathcal{C}}\frac{i}{\u{\o}_0\cdot\u{\nu}}
\left[g_{\u{\nu}}+(m\u{\Omega}\cdot\Pa Z)_{\u{\nu}}+m_{\u{\nu}}a\Omega^2\right] 
e^{i\u{\nu}\cdot\u{\phi}}.
\end{eqnarray*}
For each iteration, we express the Hamiltonian in the new action and angle 
variables. This is done recursively, following Ref.~\cite{cgjk}.
This KAM-type iteration adds terms of order $\varepsilon$
to $m$ and to the resonant part of $f$ and $g$.\\
We iterate this procedure
in order to reduce completely the nonresonant part of $f$ and $g$:
$$
H'=H\circ U_H, \, \mbox{ where } U_H=U_{S_1}\circ U_{S_2} \circ \cdots
U_{S_n}\circ \cdots,
$$
where $\I f=\I g=0$.
We notice that step (4) does not change $\u{\o}_0$ and $\u{\Omega}$
[as opposed to steps (1), (2), and (3)]. \\
The transformation $U_H$ is rigorously defined for a sufficiently
small perturbation, consisting of $f$ and $g$ (see Ref.~\cite{koch}), 
but the convergence in the whole domain of existence
of the torus is a conjecture based on numerical observations.\\
In summary the renormalization transformation acts as follows: First, some
of the resonant modes are turned into nonresonant ones (by a rescaling of 
phase space). Then a KAM-type iteration eliminates these nonresonant modes,
while slightly changing the resonant ones.

\section{Comments}
\setcounter{equation}{2}

The numerical implementation of the renormalization scheme for a 
given frequency $\o$ shows that there are two main domains separated
by a surface: one where the iteration converges
to $f=g=0$ and the other where it diverges to 
infinity.\\
The conjecture is that the boundary of the domain of convergence of the 
transformation, $\partial D$, coincides (at least locally, not too far
away from the nontrivial fixed set) with
the critical surface (where the torus
is critical). This is by no means trivial since the transformation 
is based on properties that are valid for small $f$ and $g$. However,
for a one-parameter family, the numerical evidence of the coincidence
between the critical coupling (where the torus breaks up, determined by
Greene's criterion) and the value of the parameter where the iteration starts
to diverge, indicates that we can expect $\partial D$ to coincide
with the critical surface, 
at least in a large region containing the nontrivial fixed set.\\
In this section, 
we analyze the renormalization flow on the basis of the continued
fraction expansion of the frequency.\\

{\em Quadratic irrational frequencies}-- We start by analyzing
the effect on $\o$ and $\alpha$ of $s$ renormalization steps.
Denote by $\{b_j\}$ the continued fraction expansion of $\alpha$:
$\alpha=[b_0,b_1,\ldots]$. The renormalization $\mathcal{R}_{a_{s-1}}
\mathcal{R}_{a_{s-2}}\cdots\mathcal{R}_{a_0}$ changes $\o=[a_0,a_1,\ldots]$
into $[a_s,a_{s+1},\ldots]$, and $\alpha$ into $[a_{s-1},a_{s-2},\ldots,a_0,
b_0,b_1,\ldots]$.\\
If $\o$ has a periodic continued fraction expansion
of period $s$, i.e., $\o=[(a_1,\ldots,a_s)_\infty]$,
one expect to have a nontrivial fixed point on the critical surface 
of the renormalization transformation in which one step is
defined by the composition $\mathcal{R}_{a_s}
\mathcal{R}_{a_{s-1}}\cdots\mathcal{R}_{a_1}$. We notice that
$\alpha$ converges to $[(a_s,\ldots,a_1)_\infty]$.
Therefore $\u{\Omega}$ converges to the unstable eigenvector of the
matrix $N_{a_s}\cdots N_{a_1}$ (the stable eigenvector of this matrix is
$\u{\o}_0$).\\
The nontrivial fixed point associated
with $\o$ defines a universality class that we characterize with 
critical exponents such as the total rescaling of phase space (product
of the $s$ rescalings), and the unstable eigenvalue of the linearized map
around the fixed point.\\
Two frequencies $\o_1$ and $\o_2$ having the same periodic tail
(and different first entries) in their continued fraction expansions,
belong to the same universality class. The initial integers in the continued
fraction expansion are irrelevant.\\
Associated with this nontrivial fixed point, we also have  nontrivial
fixed sets related to the nontrivial fixed point by symmetries. Therefore 
these hyperbolic sets belong to the same universality class. According
to Ref.\ \cite{chandre}, these sets are given by the intertwining 
relation
$$
\mathcal{R}_{a_0}\circ \mathcal{T}_{\u{\theta}}=\mathcal{T}_{N_{a_0}\u{\theta}}
\circ \mathcal{R}_{a_0},
$$
where $\mathcal{T}_{\u{\theta}}$ is defined as the translation 
$(\mathcal{T}_{\u{\theta}} f)(\u{\varphi})=f(\u{\varphi}+\u{\theta})$.
Applying the relation
$$\mathcal{R}_{a_s}\circ \cdots \circ \mathcal{R}_{a_1}
\circ \mathcal{T}_{\u{\theta}}=\mathcal{T}_{N_{a_s}\cdots N_{a_1}\u{\theta}}
\circ \mathcal{R}_{a_s}\circ \cdots \circ \mathcal{R}_{a_1}$$
to the fixed point $H_*(\u{\phi})$, we have
$$
\mathcal{R}_{a_s}\circ \cdots \circ \mathcal{R}_{a_1} H_*(\u{\phi}
+\u{\theta})=H_*(\u{\phi}+N_{a_s}\cdots N_{a_1}\u{\theta}).
$$
The map 
$$
\u{\theta} \mapsto N_{a_s}\cdots N_{a_1}\u{\theta} \mbox{ mod } 2\pi,
$$
gives the nature of the orbits to which the transformation converges.\\
For instance, for the golden mean case, one has a nontrivial fixed point
and a nontrivial 3-cycle in the space of even Hamiltonians. For Hamiltonians
without parity restriction, the nontrivial fixed sets can be labeled
by the orbits of the Anosov map 
$N=\left(\d\begin{array}{cc} 1 & 1\\ 1 & 0 \end{array}\right)$
($N^2$ is Arnold's cat map).
For $\o=[(1,2)_\infty]$, one has a nontrivial 2-cycle and two fixed points
in the space of even Hamiltonians.\\
In the perturbative regime, there exists a geometrical accumulation 
of a certain sequence of periodic orbits with a ``trivial'' ratio
(trivial in the sense that it is explicit, see comment in the previous section).
The fact that it also happens in the
critical case, but with a nontrivial ratio
(which is the unstable eigenvalue of the linearized map around the
nontrivial fixed point), implies universal self-similar
properties of the critical torus~\cite{kadanoff,shenker}.\\
As already mentioned in Ref.~\cite{lanford}, $\mathcal{R}_{a_i}$ becomes
more and more singular as $a_i$ increases. Therefore the picture we
present is valid only for bounded $a_i$. For large $a_i$, there is, to our
knowledge, no description of the renormalization flow and its properties, 
even from a numerical point of view.\\

{\em Nonquadratic irrational frequencies}-- In that case, we cannot expect 
any geometrical accumulation of periodic orbits even if the Hamiltonian
is close to integrable (perturbative regime). The renormalization has no
fixed point, but instead its flow is related to a chaotic trajectory of the 
Gauss map 
$$
\o \mapsto 1/\o - [1/\o],$$
where $[\, ]$ denotes the integer part. We notice that $\alpha$ follows
the inverse of this trajectory.\\
For a typical $\o\in ]0,1[$, it is numerically impossible to figure out
what could be the critical set, as one only works with finite precision
(and thus a finite number of entries in the continued fraction, determined
by iterating the Gauss map). \\
Instead, one should give a sufficient number of entries in the continued
fraction expansion, and then iterate the renormalization transformation
on the critical surface. These entries should be bounded in order to avoid
singularities mentioned above.\\
On the critical surface, one can conjecture the
existence of a critical strange attractor. But
these hyperbolic sets should be conceptually different from the
sets found in the quadratic irrational case (related to the nontrivial fixed
point by symmetries): there is a continuous distribution of critical 
exponents (e.g., rescalings). The mean-rescaling (geometric mean value)
and the largest Lyapunov exponent characterize the universal class associated
with $\o$.\\
In order to have information about a typical $\o\in ]0,1[$, an ergodic
renormalization for random continued fractions has been proposed
\cite{ostlund,satija}. It consists in determining universal parameters
by averaging over a large number of random continued fractions, constructed 
following a given probability distribution for the coefficients.\\
The conjecture is that ergodic renormalization trajectories converge
to a strange attractor (which contains all the nontrivial fixed sets obtained
for quadratic irrationals). Lyapunov exponents and other quantities such
as the mean-rescaling are universal. Related ideas have been proposed
for circle maps in Refs.~\cite{lanford,lanford2}.\\

{\em Extension to higher dimensional systems\/}-- We have seen that the 
existence of a nontrivial fixed point was based on a {\em geometrical}
accumulation of a certain sequence of periodic orbits. These were given
by the truncations of the continued fraction expansion (for quadratic 
irrationals). For three degrees of freedom, we lack of a theory that
generalizes the continued fractions. However, we can choose a torus such
that it is a geometrical accumulation of periodics orbits, i.e., the
sequence of resonances is generated by a {\em single} matrix
(as it was the case for the golden mean). It allows us
to define a renormalization-group transformation with a {\em fixed} frequency 
vector. Even in that case, preliminary studies~\cite{3d} suggest that one 
can expect a strange attractor instead of a nontrivial fixed point. 
This feature
depends strongly on the spectrum of the matrix which generates the
resonances.

\section{Analyticity properties of the nontrivial fixed point}
\setcounter{equation}{3}

As the renormalization transformation described in the previous sections
does not reduce the nonresonant part of $m$ (it only reduces
 $g$ and $f$) in order to remain quadratic in the actions, the shift 
of the Fourier modes [step (1)] reduces the analyticity of $m$ at each step. 
As a consequence, the nontrivial fixed point is expected to have a 
nonanalytic $m$. 
The numerical implementation of the transformation gives accurate computation
of the critical exponents characterizing the universality class, despite
the fact that
$m$ is nonanalytic at the nontrivial fixed point.
At present, this fact is not well understood.\\
A method that can yield an analytic nontrivial fixed point
is to eliminate the nonresonant part of $m$ together 
with the one of $g$ and $f$~\cite{abad}. 
The main drawback of this method is that one
cannot remain quadratic in the actions. This drastically 
complicates the numerical implementation of the transformation as one
needs to work with $N$ scalar functions instead of 3, $N$ denoting the
numerical truncation of the Taylor series in the actions.

\section*{Acknowledgments}
We acknowledge very
useful discussions with G. Benfatto, G. Gallavotti, H. Koch,
and R.S. MacKay.
Support from the EC Contract No.\ ERBCHRXCT94-0460 for the project
``Stability and Universality in Classical Mechanics''
and from  the Conseil R\'egional de Bourgogne is acknowledged.

\end{document}